\documentclass[twocolumn]{aastex631}
\usepackage{array, makecell}
\usepackage{multirow}
\usepackage{cancel}
\usepackage{comment}

\begin{document}

\title{Intriguing Plasma Composition Pattern in a Solar Active Region: a Result of Non-Resonant Alfv\'en Waves?
}

\correspondingauthor{Teodora Mihailescu}
\email{teodora.mihailescu.19@ucl.ac.uk}

\author[0000-0001-8055-0472]{Teodora Mihailescu}
\affil{Mullard Space Science Laboratory, University College London, Holmbury St Mary, Dorking, Surrey, RH5 6NT, UK}

\author[0000-0002-2189-9313]{David H. Brooks}
\affil{College of Science, George Mason University, 4400 University Drive, Fairfax, VA 22030, USA}

\author[0000-0002-3362-7040]{J. Martin Laming}
\affil{Space Science Division, Naval Research Laboratory, Code 7684, Washington, DC 20375, USA}

\author[0000-0002-0665-2355]{Deborah Baker}
\affil{Mullard Space Science Laboratory, University College London, Holmbury St Mary, Dorking, Surrey, RH5 6NT, UK}

\author[0000-0002-0053-4876]{Lucie M. Green}
\affil{Mullard Space Science Laboratory, University College London, Holmbury St Mary, Dorking, Surrey, RH5 6NT, UK}

\author[0000-0001-7927-9291]{Alexander W. James}
\affil{Mullard Space Science Laboratory, University College London, Holmbury St Mary, Dorking, Surrey, RH5 6NT, UK}

\author[0000-0003-3137-0277]{David M. Long}
\affil{Astrophysics Research Centre, Queen’s University Belfast, University Road, Belfast, BT7 1NN, Northern Ireland, UK}

\author[0000-0002-2943-5978]{Lidia van Driel-Gesztelyi}
\affil{Mullard Space Science Laboratory, University College London, Holmbury St Mary, Dorking, Surrey, RH5 6NT, UK}
\affil{LESIA, Observatoire de Paris, Université PSL, CNRS, Sorbonne Université, Univ. Paris Diderot, 5 place Jules Janssen, F-92195 Meudon, France}
\affil{Konkoly Observatory, Research Centre for Astronomy and Earth Sciences, Konkoly Thege út 15-17., H-1121, Budapest, Hungary}

\author[0000-0002-5365-7546]{Marco Stangalini}
\affil{ASI Italian Space Agency, Via del Politecnico, s.n.c I-00133—Roma, Italy}



\begin{abstract}

The plasma composition of the solar corona is different from that of the solar photosphere. Elements that have a low first ionisation potential (FIP) are preferentially transported to the corona and, therefore, show enhanced abundances in the corona compared to the photosphere. The level of enhancement is measured using the FIP bias parameter. In this work, we use data from the EUV Imaging Spectrometer (EIS) on Hinode to study the plasma composition in an active region following an episode of significant new flux emergence into the pre-existing magnetic environment of the active region. We use two FIP bias diagnostics: Si X 258.375~\AA/S X 264.233~\AA\ (temperature of approximately 1.5 MK) and Ca XIV 193.874~\AA/Ar XIV 194.396~\AA\ (temperature of approximately 4 MK). We observe slightly higher FIP bias values with the Ca/Ar diagnostic than Si/S in the newly emerging loops, and this pattern is much stronger in the preexisting loops (those that had been formed before the flux emergence). This result can be interpreted in the context of the ponderomotive force model, which proposes that the plasma fractionation is generally driven by Alfv\'en waves. Model simulations predict this difference between diagnostics using simple assumptions about the wave properties, particularly that the fractionation is driven by resonant/non-resonant waves in the emerging/preexisting loops. We propose that this results in the different fractionation patterns observed in these two sets of loops.

\end{abstract}



\section{Introduction} \label{Introduction}
One of the major open questions in solar physics is why the elemental composition of some regions in the solar corona is different to that of the underlying photosphere. The relative abundance of different elements is spatially homogenous in the photosphere \citep{asplund_chemical_2009}, but varies in the corona, with a strong dependence on the first ionization potential (FIP) of the element \citep{meyer_baseline_1985}. The abundances of elements with a low FIP ($<$10~eV) are often enhanced in the corona, while the abundances of high FIP elements ($>$10~eV) appear to be unchanged. This is called the FIP effect. The degree of enhancement of an element is calculated using the FIP bias parameter. In the extreme ultraviolet (EUV), this is measured as the abundance ratio of a low-FIP element to a high-FIP element, relative to the same photospheric ratio. Observed FIP bias values typically vary between 1 and 4 \citep[e.g.][]{baker_plasma_2013, del_zanna_multi-thermal_2013} but higher values that go up to 8 have also been reported \citep[e.g.][]{widing_rate_2001}.

Various processes have been proposed to be responsible for this effect, such as diffusion or inefficient Coulomb drag \citep[e.g.][]{von_steiger_supply_1989, marsch_element_1995, pucci_elemental_2010, bo_effect_2013}, thermoelectric driving \citep{antiochos_physics_1994}, chromospheric reconnection \citep{arge_modelling_1998} or ion cyclotron wave heating \citep{schwadron_elemental_1999} . Among the many candidates, a collisionless wave-particle mechanism based on the ponderomotive force \citep{laming_unified_2004, laming_fip_2015} appears to be able to describe this phenomenon more realistically than previously suggested mechanisms. The ponderomotive force arises as the reaction of the plasma to the refraction of Alfv\'en waves in the chromosphere and only acts on charged particles. In the chromosphere, low-FIP elements are predominantly ionized \citep[ionization fraction $>$99\%; see][and references therein]{laming_fip_2015}, while high-FIP elements are mostly neutral. Therefore, the ponderomotive force separates the low-FIP ions from the high-FIP neutrals by preferentially transporting them upwards to the corona. The result can then be observed in the corona as enhanced abundances of low-FIP elements in the corona compared to the photosphere. This is interesting because it places the fractionation closer to the beginning of the chain of events that lead to coronal heating, rather than being at the endpoint of the thermalization of that energy as would be the case for some of the diffusion based mechanisms. Numerical simulations by \citet{dahlburg_ponderomotive_2016} support the presence of the ponderomotive acceleration in solar coronal loops, with the appropriate magnitude and direction, and suggest it is a “by-product” of coronal heating. Recent work by \citet{baker_alfvenic_2021, stangalini_spectropolarimetric_2021, murabito_investigating_2021} found magnetic fluctuations in the chromosphere being magnetically connected to regions of high FIP bias in the corona which supports this theoretical model. More recent numerical simulations by \citet{reville_investigating_2021} using a shell turbulence model found that, under the assumption that turbulence is the main driver of coronal heating and solar wind acceleration, a ponderomotive force can appear in the chromosphere and the transition region, and can be strong enough to create the FIP effect. \citet{martinez-sykora_impact_2023} use a combination of IRIS observations and a 2.5D radiative magnetohydrodynamics (MHD) model of the solar atmosphere to investigate the multifluid effects on the ponderomotive force associated with Alfv\'en waves.

The strongest FIP bias is typically observed in active regions. There are large variations in the measured FIP bias among active regions \citep{brooks_full-sun_2015, mihailescu_what_2022} and the overall FIP bias values observed in an active region also vary with time as the active region goes through the different stages that make up its lifetime. Previous studies found that emerging flux carries plasma with photospheric composition, i.e., FIP bias of approximately 1, at first \citep{widing_rate_2001}. Then it increases with time in the emergence phase \citep{widing_rate_2001} and early decay phase \citep{baker_coronal_2018}, suggesting that the higher level of magnetic activity observed during the emergence phase is linked to the processes that drive the FIP effect. After an active region enters its decay phase, the FIP bias starts to decrease \citep{baker_fip_2015, ko_correlation_2016} until it reaches the FIP bias of the surrounding quiet Sun \citep{ko_correlation_2016}. 

In addition, FIP bias values within an active region show a broad distribution \citep{mihailescu_what_2022}, indicating that processes acting on sub-active region scales can influence the FIP bias in different ways. For example, \citet{baker_plasma_2013} found that the FIP bias is highest at the loop footpoints and shows a mild enhancement along some of the active region loops. The same study also found that photospheric reconnection that manifests as photospheric flux cancellation and subsequent formation of a flux rope leads to FIP bias values closer to 1. This makes plasma composition a powerful tool for obtaining insight into the magnetic configuration and formation history of solar structures \citep[see e.g.,][]{fletcher_brightenings_2001, james_-disc_2017, baker_evolution_2022}.

The measured FIP bias values can also vary depending on the diagnostic used to measure it. While elements are broadly categorised into low-FIP and high-FIP, different low-FIP elements can show different levels of enhancement, and high-FIP elements can in some instances show enhancement too. The clearest example is S (FIP = 10.36 eV) which sits close to the boundary between low-FIP and high-FIP elements. In some instances it shows no or little enhancement (i.e., it behaves like a high-FIP element), while in others it shows significant enhancement \citep[i.e., it behaves like a low-FIP element, see e.g., the coronal hole measurements of][]{brooks_establishing_2011}. In addition, in the EUV, FIP bias diagnostics also have an associated temperature given by the contribution functions of the lines used. For example, the Si {\scriptsize X} 258.38 \AA/S {\scriptsize X} 264.22 \AA\ diagnostic captures cooler coronal plasma as the formation temperature of the lines involved is around 1.5~MK, while the Ca {\scriptsize XIV} 193.87 \AA/Ar {\scriptsize XIV} 194.40 \AA\ diagnostic captures hotter coronal plasma since the lines involved form at around 4~MK. Therefore, differences in FIP bias values measured with different diagnostics can be due to either the elements themselves behaving differently or the diagnostic probing plasma at different temperatures. \citet{ko_correlation_2016} found a high correlation (correlation coefficient varying between 0.76 and 0.91 depending on the region selected) between the FIP bias measured with Si {\scriptsize X} 258.38 \AA/S {\scriptsize X} 264.22 \AA\ and Fe {\scriptsize XII} 195.12 \AA/S {\scriptsize X} 264.22 \AA\ ($\text{log}(\text{T}_{\text{MAX}})=6.2$) in a decaying active region. This indicates that Fe and Si show similar FIP enhancement relative to S (note that S is the high-FIP element in both diagnostics). \citet{to_evolution_2021} found more significant differences between FIP bias diagnostics when analysing the Si {\scriptsize X} 258.38 \AA/S {\scriptsize X} 264.22 \AA\ compared to Ca {\scriptsize XIV} 193.87 \AA/Ar {\scriptsize XIV} 194.40 \AA\ FIP bias values in a small flare. This study suggests that the elements considered to be high-FIP could be behaving differently, i.e., S could be acting like a low-FIP element, as previously suggested by \citet{laming_element_2019}. However, the conditions under which this phenomenon can happen are not yet fully understood.

This work investigates strikingly different fractionation patterns in different parts of an active region, again involving variations in the behavior of S. We use a combination of two FIP bias diagnostics characterising the two fractionation patterns and ponderomotive force model simulations to further investigate mechanisms that could be responsible for the observed differences. Section \ref{Evolution} presents the photospheric and coronal evolution of the active region in the time leading up to the plasma composition observations. Section \ref{Composition Measurements} describes the composition observations and the method for obtaining the FIP bias measurements. Section \ref{The Ponderomotive Force Model} presents the ponderomotive force model \citep{laming_unified_2004, laming_fip_2015} simulation results and how they can contribute to understanding the plasma composition observations. Finally, Section \ref{Discussion} provides a summary and discussion of the results in this study. 

\section{Evolution of AR 12665}
\label{Evolution}
\subsection{White Light Continuum and Photospheric Magnetic Field Evolution}
\begin{figure}
    \centering
    \includegraphics[width=0.5\textwidth]{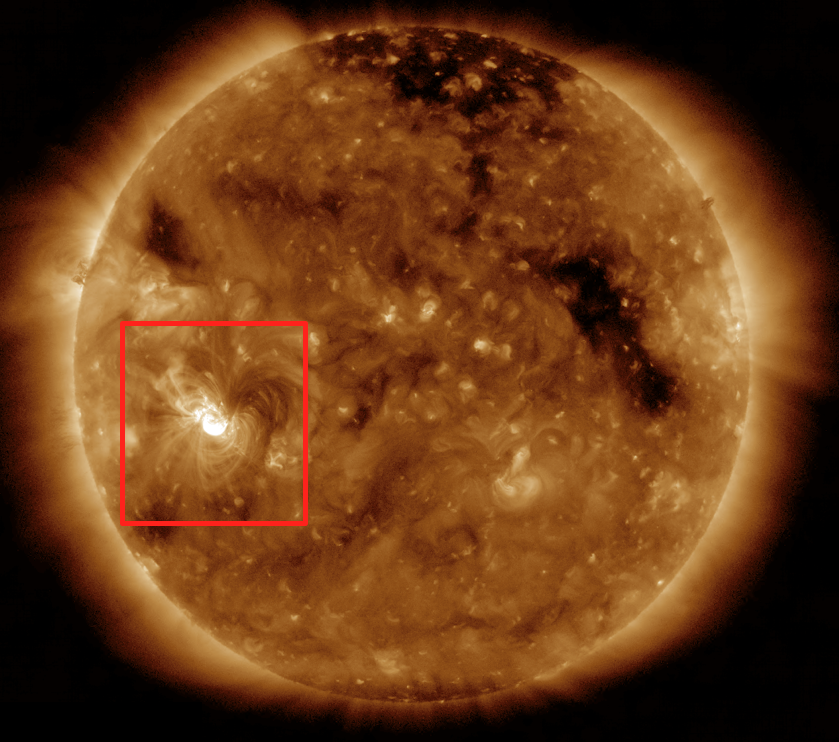}
    \caption{SDO AIA 193 \AA\ context image of target AR 12665 on 2017 July 9 at 03:59 UT.}
    \label{AR12665}
\end{figure}
\begin{figure*}
    \centering
    \includegraphics[width=1.0\textwidth]{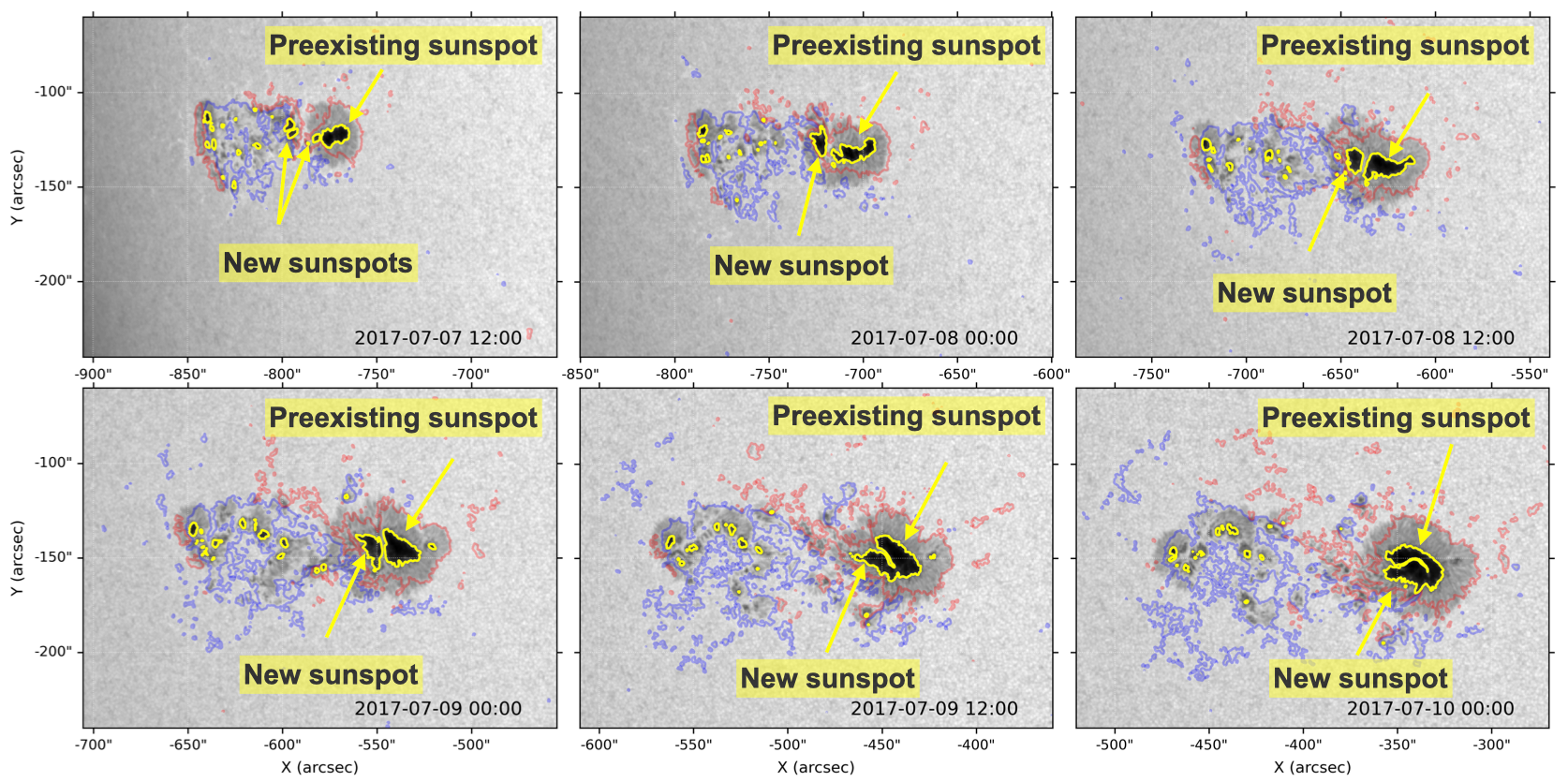}
    \caption{SDO HMI Continuum emission and photospheric magnetic field evolution of AR 12665, every twelve hours, prior to and during the Hinode/EIS scans. Red (blue) contours represent areas of HMI line of sight photospheric magnetic field strength above (below) 200 G (-200 G). Yellow contours represent values below 25,000 ct/s in the continuum emission, indicating the location of the sunspot umbrae.}
    \label{Continuum_evolution}
\end{figure*}

NOAA AR 12665 (see Figure \ref{AR12665}) was first observed by the Solar Dynamics Observatory \citep[SDO;][]{pesnell_solar_2012} Helioseismic and Magnetic Imager \citep[HMI;][]{scherrer_helioseismic_2012,schou_design_2012} at the eastern limb on 2017 July 5, with sunspots already present in both polarities. A new episode of significant flux emergence had begun between the leading and following polarities just as the AR region rotated into Earth view. By July 8, the emerging flux formed a new positive sunspot, which started moving towards (see top panels in Figure \ref{Continuum_evolution}) and then orbiting the preexisting leading sunspot in a counterclockwise direction (see bottom panels in Figure \ref{Continuum_evolution}). They eventually collided around 12:00 UT on July 9 and became one sunspot consisting of two umbrae separated by a light bridge within one common penumbra. Approximately one day later the light bridge disappeared (not pictured). This orbiting motion lasted for multiple days and was studied in detail by \citet{james_new_2020}. For the first approximately 1.5 days, the light bridge constantly separated the new from the preexisting positive sunspot umbrae. This enabled us to track the evolution of the newer and older parts of the active region separately.

\subsection{Coronal Evolution}
\begin{figure*}
    \centering
    \includegraphics[width=0.8\textwidth]{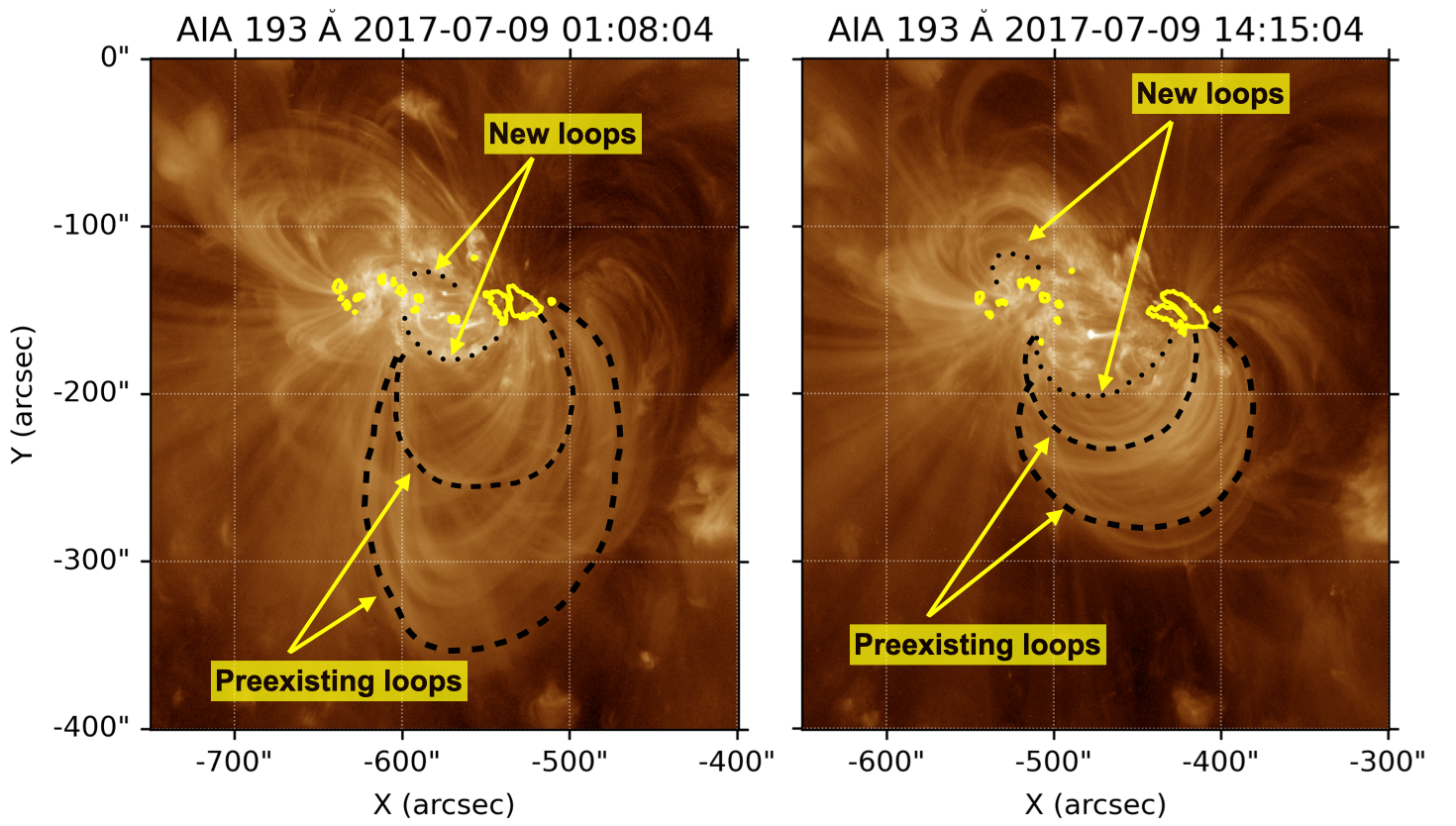}
    \caption{SDO AIA 193 \AA\ passband images at the times matching the middle time of the EIS raster scans. Yellow contours represent values below 25,000 ct/s in the continuum emission, indicating the location of the sunspots. Black dotted (dashed) lines indicate representative examples of loops belonging to the new (preexisting) loop populations.}
    \label{AIA193zoom}
\end{figure*}

In the EUV, images from the SDO Atmospheric Imaging Assembly \citep[AIA;][]{lemen_atmospheric_2012} showed two main loop populations, namely the new loops and the preexisting loops. The new loops had been formed recently by the flux emergence and were rooted in the newly formed positive sunspot (see Figure \ref{Continuum_evolution}). They were bright, relatively small hot loops in the core of the active region (see Figure \ref{AIA193zoom}). The preexisting loops had been part of the active region since before the flux emergence and were rooted in the preexisting positive sunspot (see Figure \ref{Continuum_evolution}). They were fainter, high-lying warm loops located in the southern part of the active region (see Figure \ref{AIA193zoom}). 

In the few days running up to the EIS scans on July 9 at 01:08 UT and 14:15 UT, the active region showed modest activity, with only one C1.0 flare on July 7 at 13:37 UT and one C3.4 flare on July 8 at  23:50 UT. On July 9, however, the flaring activity becomes more intense, with one M1.3 flare at 03:09 UT and four C-class flares at 06:15 UT, 07:28 UT, 08:55 UT and 11:44 UT being observed at the boundary between the new and the preexisting loop populations. 

\section{Plasma Composition}
\label{Composition Measurements}

\subsection{Hinode EIS Observations}
The FIP bias was calculated using observations from the EUV Imaging Spectrometer \citep[EIS;][]{culhane_euv_2007} on Hinode \citep{kosugi_hinode_2007}. The EIS dataset contains 6 scans of the active region over a period of approximately 13 hours. In this work, we analyse the first and the last scans in this series. Details of the EIS scans used are provided in Table \ref{EIS_studies}.
\begin{deluxetable}{ll}[t]
\tabletypesize{\footnotesize}
\centering
\tablecolumns{2}
\tablehead{\colhead{EIS Study Details}}
\tablecaption{Summary of Hinode/EIS study details and emission lines used for creating the FIP bias measurements.} \label{EIS_studies}
\startdata
Raster middle times &\makecell[l]{09/07/2017 01:08 \\ 09/07/2017 14:15}\\
Study acronym & DHB\_007\\
Study number & 544 \\
Field of view & 492'' $\times$ 512''\\
Rastering & 2'' slit, 123 positions, 4'' coarse step\\
Exposure time & 30s\\
Total raster time & 1h 1m 30s\\
\makecell[l]{Reference spectral \\ window} & Fe {\scriptsize XII} 195.12 \AA{} \\ 
DEM lines & \makecell[l]{Fe {\scriptsize VIII} 185.213 \AA, Fe {\scriptsize VIII} 186.601 \AA, \\ Fe {\scriptsize IX} 188.497 \AA, Fe {\scriptsize IX} 197.862 \AA, \\ Fe {\scriptsize X} 184.536 \AA, Fe {\scriptsize XI} 188.216 \AA, \\ Fe {\scriptsize XI} 188. 299 \AA, Fe {\scriptsize XII} 192.394 \AA, \\ Fe {\scriptsize XII} 195.119 \AA, Fe {\scriptsize XIII} 202.044 \AA, \\ Fe {\scriptsize XIII} 203.826 \AA, Fe {\scriptsize XIV} 264.787 \AA, \\ Fe {\scriptsize XIV} 270.519 \AA, Fe {\scriptsize XV} 284.16 \AA, \\ Fe {\scriptsize XVI} 262.984 \AA, Ca {\scriptsize XIV} 193.874 \AA, \\ Ca {\scriptsize XV} 200.972 \AA.}\\
Density diagnostic lines & \makecell[l]{Fe {\scriptsize XIII} 202.04 \AA, Fe {\scriptsize XIII} 203.82 \AA}\\
Line ratio lines & \makecell[l]{Si {\scriptsize X} 258.38 \AA, S {\scriptsize X} 264.22 \AA, \\ Ca {\scriptsize XIV} 193.87 \AA, Ar {\scriptsize XIV} 194.40 \AA}\\
\enddata
\end{deluxetable}

\subsection{Method}
\begin{figure*}
    \centering
    \includegraphics[width=0.94\textwidth]{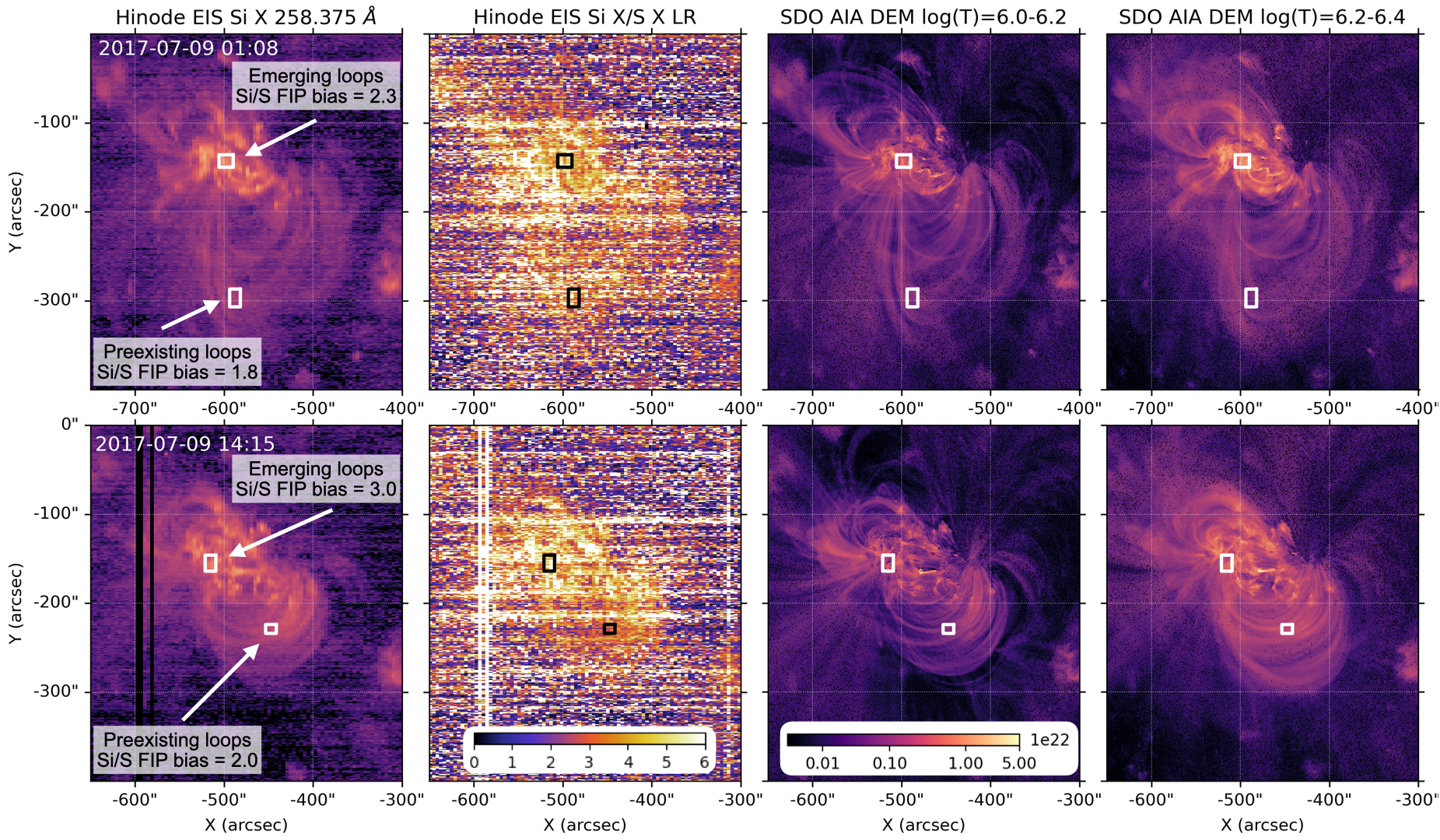}
    \caption{Si {\scriptsize X}/S {\scriptsize X} results. From left to right: Hinode EIS Si {\scriptsize X} 258.38 \AA\ intensity, Hinode EIS Si {\scriptsize X} 258.38 \AA/S {\scriptsize X} 264.22 \AA\ line ratio, SDO AIA DEM in the log(T)=6.0 - 6.2 and the log(T)=6.2 - 6.4 temperature bins \citep[computed using the method developed by][]{hannah_differential_2012, hannah_multi-thermal_2013}. The boxes indicate the locations of the macropixels for this diagnostic.}
    \label{SiS_results}
    \includegraphics[width=0.94\textwidth]{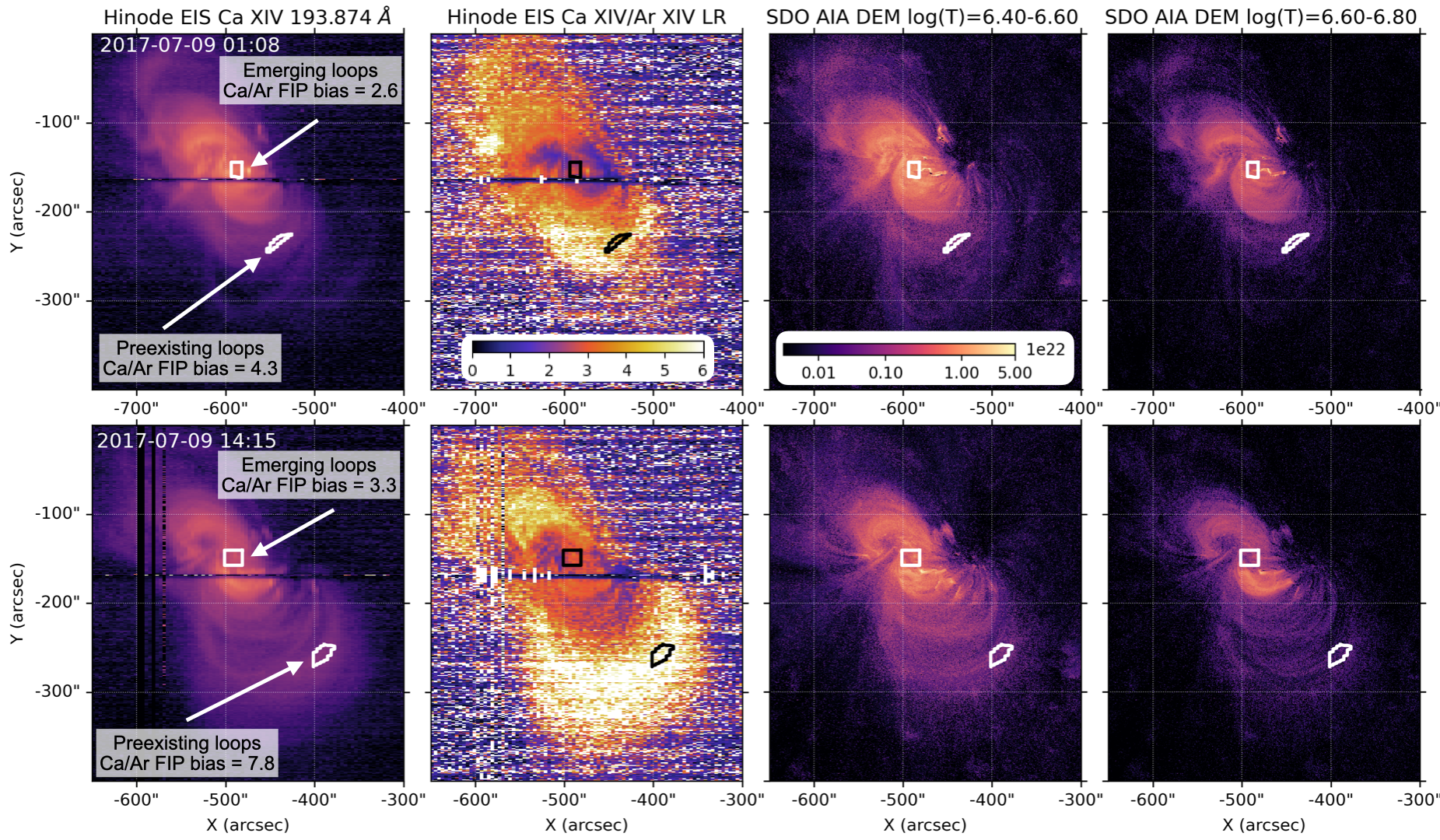}
    \caption{Ca {\scriptsize XIV}/Ar {\scriptsize XIV} results. From left to right: Hinode EIS Ca {\scriptsize XIV} 193.87 \AA\ intensity, Hinode EIS Ca {\scriptsize XIV} 193.87 \AA/Ar {\scriptsize XIV} 194.40 \AA\ line ratio, SDO AIA DEM in the log(T)=6.4 - 6.6 and the log(T)=6.6 - 6.8 temperature bins \citep[computed using the method developped by][]{hannah_differential_2012, hannah_multi-thermal_2013}. The boxes indicate the locations of the macropixels for this diagnostic.}
    \label{CaAr_results}
\end{figure*}

We used two line pair diagnostics: Si {\scriptsize X} 258.38 \AA\ (low FIP, FIP = 8.25 eV) \&  S {\scriptsize X} 264.22 \AA\ (high FIP, FIP = 10.36 eV) and Ca {\scriptsize XIV} 193.87 \AA\ (low FIP, FIP = 6.11 eV) \& Ar {\scriptsize XIV} 194.40 \AA\ (high FIP, FIP = 15.76 eV). According to the CHIANTI database \citep{dere_chianti_1997} Version 10 \citep{del_zanna_chiantiatomic_2021}, the theoretical formation temperatures for the two diagnostics are different: the Si {\scriptsize X} 258.38 \AA\ and S {\scriptsize X} 264.22 \AA\ lines have a formation temperature of $\text{log}(\text{T}_{\text{MAX}}) = 6.2$, while the Ca {\scriptsize XIV} 193.87 \AA\ and Ar {\scriptsize XIV} 194.40 \AA\ form at a temperature of $\text{log}(\text{T}_{\text{MAX}}) = 6.7$ and $\text{log}(\text{T}_{\text{MAX}}) = 6.6$ respectively. The lines involved in the two diagnostics were fitted using the Python EISPAC software \citep{weberg_eispac_2023} and line ratios for each diagnostic were calculated in every pixel in the EIS rasters  to obtain an approximation of the FIP bias (see second column in Figures \ref{SiS_results} and \ref{CaAr_results}). 

Line ratios, however, are sensitive to temperature and density effects, so they only provide context. Corrected FIP bias measurements were also calculated in a few key locations. In these locations, spectra were averaged over multiple pixels (creating a macropixel) for all the lines included in the calculation. The method used for the FIP bias calculation in each of these macropixels uses a differential emission measure (DEM) to correct the temperature effects and a density analysis to correct for density effects. The DEM was derived using a series of Fe lines supplemented with a couple of Ca lines for additional high temperature constraints (see Table \ref{EIS_studies}) and the density was calculated using the Fe {\scriptsize XIII} 202.04 \AA/Fe {\scriptsize XIII} 203.82 \AA\ diagnostic. A Markov-Chain Monte Carlo (MCMC) algorithm in the PintOfAle \citep{kashyap_markovchain_1998,kashyap_pintofale_2000} SolarSoft \citep{freeland_data_1998} package was used to compute the DEM distribution, and the CHIANTI database \citep{dere_chianti_1997} version 10 \citep{del_zanna_chiantiatomic_2021} to compute the contribution functions (G(T,n)) for each of the spectral lines involved in the diagnostic. We used the photospheric abundances of \cite{scott_elemental_2015} and \cite{scott_elemental_2015-1}. Note that using different sets of photospheric abundances could result in slightly different FIP bias measurements. This method is described in detail by \citet{brooks_establishing_2011, brooks_full-sun_2015}.

There are 8 such locations in total. For each of the 2 rasters, 4 macropixels were selected: one for each loop population and each diagnostic (see first panel of Figures \ref{SiS_results} and \ref{CaAr_results}). This was done to obtain representative FIP bias values for each of the loop populations in both diagnostics. Slightly different macropixels were chosen for the two diagnostics. This is because the formation temperatures for the lines involved in the two diagnostics are different. For each diagnostic, the macropixels were chosen such that emission in the lines involved in the diagnostic is maximised.

\subsection {Results}
The FIP bias values are summarised in Table \ref{FIP_results}. In the emerging loops, the Si/S FIP bias increases with time from 2.3 to 3.0 and the Ca/Ar FIP bias from 2.6 to 3.3. This result agrees with previous studies which found increasing FIP bias in the emergence phase of an active region \citep{widing_rate_2001, baker_coronal_2018}. While the increase is observed in both diagnostics, the Ca/Ar values are slightly higher than the Si/S ones \citep[albeit close to the 0.3 uncertainty in the measurements previously estimated by][]{brooks_solar_2017}.

In the preexisting loops, the FIP bias behaviour is different. The Si/S FIP bias changes slightly from 1.8 to 2.0, but this is within the 0.3 error limit. The Ca/Ar FIP bias shows an increase from 4.3 to 7.8 and, more importantly, shows consistently high values. This is very interesting because these values are significantly higher than the Si/S FIP bias values in the same loop population. To check that these high values are indeed representative of the entire preexisting loops population, we examine the line ratio maps (see second column of Figure \ref{CaAr_results}). The line ratio maps show high values everywhere in the preexisting loops population, suggesting that the high FIP bias values are not isolated to the location of the macropixels that were chosen for the FIP bias calculation. Of course, we must be cautious when analysing line ratio maps as they are sensitive to temperature and density effects. In particular, for the Ca/Ar line ratio, significant temperature effects due to plasma above $\text{log}(T)=6.6$ must be considered \citep[see e.g.][]{feldman_can_2009, doschek_sunspots_2017, to_evolution_2021}. The DEM analysis (see third and fourth panel in Figures \ref{SiS_results}, \ref{CaAr_results}) shows that the emission in these preexisting loops is mostly cooler, with most of the emission coming from the $\text{log}(T)=6.2-6.4$ and $\text{log}(T)=6.4-6.6$ temperature bins, making it unlikely that the high values we see is a temperature effect. 

The large Ca/Ar FIP bias values observed in the preexisting loops are not common, but similarly high values have been observed before, for example, in the Ca/Ar ratio in post flare loops \citep{doschek_photospheric_2018}, in the Mg/Ne ratio in an emerging flux region \citep{young_mgne_1997, widing_rate_2001}, in the Mg/O ratio in coronal mass ejection cores \citep{landi_physical_2010}, and in various diagnostics in post coronal mass ejection current sheets \citep{ko_dynamical_2003, ciaravella_elemental_2002}. The newest aspect of the present results is the different behaviour of the two diagnostics used to probe the plasma composition in the two loop populations. While in the emerging loops, the Si/S and Ca/Ar diagnostics indicate similar FIP bias values, in the preexisting loops there is a significant difference between the lower Si/S FIP bias values and higher Ca/Ar FIP bias values. This raises the question of whether the mechanism driving the FIP effect has different characteristics in the two loop populations, so this possibility was explored further using simulations from the ponderomotive force model.

\begin{deluxetable}{lcccc}[t]
\tabletypesize{\footnotesize}
\centering
\tablecolumns{5}
\tablehead{\multicolumn{1}{c}{Raster Time} & \multicolumn{4}{c}{FIP Bias}}
\tablecaption{Hinode/EIS FIP bias results summary} \label{FIP_results}
\startdata
 &\multicolumn{2}{c}{Emerging Loops}&\multicolumn{2}{c}{Preexisting Loops}\\
 & Si/S & Ca/Ar & Si/S & Ca/Ar \\
 2017 July 9 01:08 UT & 2.3 & 2.6 & 1.8 & 4.3 \\
 2017 July 9 14:15 UT & 3.0 & 3.3 & 2.0 & 7.8 \\
\enddata
\end{deluxetable}

\section{The Ponderomotive Force Model}
\label{The Ponderomotive Force Model}
The ponderomotive force model \citep{laming_unified_2004, laming_fip_2015} is a 1D static model which proposes that the FIP effect is generated in the chromosphere by Alfv\'en wave activity originating in the corona. At coronal loop footpoints, refraction of these Alfv\'en waves in the high density gradient of the chromosphere (see Figure \ref{PF Model}a) generates a ponderomotive force. This ponderomotive force acts on the ionized material (i.e. mostly low-FIP elements since they have a much higher ionization fraction than the high-FIP elements, see Figure \ref{PF Model}b,c), preferentially bringing them upwards to the top of the transition region. Once plasma reaches the transition region, two things happen: 1) the temperature increases enough to ionize all elements, so low-FIP and high-FIP elements are no longer separated into ions and neutrals and 2) there is no significant density gradient anymore, so the ponderomotive force disappears. This means that the fractionation process stops at the top of the transition region, and the fractionation pattern is locked in. From here, the fractionated plasma is transported up into the corona through other mechanisms. Note that the current implementation of the ponderomotive force model uses a 1D static model chromosphere, and so it does not cover this last part of the chain.

The pattern and strength of the fractionation process depend on the height in the chromosphere at which the ponderomotive force is generated. This, in turn, is dictated by whether the Alfv\'en wave driver is in resonance with the loop or not. Resonance here means that the wave travel time from one loop footpoint to the other is an integral number of wave half-periods. Resonant waves accumulate much more wave energy in the corona than in the chromospheric footpoints, and  drive the ponderomotive force close to the top of the chromosphere (see Figure \ref{PF Model}e). This results in mild fractionation levels (see Figure \ref{PF Model}f) because of the ionized background gas. Non-resonant waves drive the ponderomotive force at lower heights (see Figure \ref{PF Model}h), which results in stronger fractionation levels (see Figure \ref{PF Model}i) because the background gas is neutral.

\subsection {Model Simulations}
\label{Model Simulations}
We use the ponderomotive force model to make predictions of the fractionation patterns in the two loop populations shown in Figures \ref{SiS_results} and \ref{CaAr_results}. We find that the simulation predictions match best with the Hinode EIS observations when assuming the driver is resonant waves in the emerging loops and non-resonant waves in the preexisting loops. This scenario is described in detail below.

We first estimate the resonant frequency of the emerging loops using the following parameters: loop length, plasma density along the loop and magnetic field strength. The loop lengths were estimated using a Potential Field Source Surface (PFSS) model \citep[IDL SolarSoft package provided by][]{schrijver_photospheric_2003} of the active region. In the PFSS model, a representative loop was selected for each loop population and its length was calculated. The plasma density along the loop was measured in the macropixel boxes shown in Figures \ref{SiS_results}, \ref{CaAr_results} using the Fe {\scriptsize XIII} 202.04 \AA/Fe {\scriptsize XIII} 203.82 \AA\ diagnostic from Hinode EIS. The photospheric magnetic field strength was measured using SDO HMI. All loop parameters are summarised in Table \ref{PF_parameters}. The resonant frequency is given approximately by
\begin{equation}
\label{eq_f_resonance}
    f_{\text{resonance}} = \frac{v_A}{2L},
\end{equation}
where L is the loop length and $v_A$ is the Alfv\'en speed calculated as:
\begin{equation}
\label{eq_alfven_speed}
v_A = \frac{B}{\sqrt{4\pi\rho}},  
\end{equation}
where B is the coronal magnetic field strength, and $\rho$ is the loop density. We construct loop models matching the parameters in Table \ref{PF_parameters}, where the 0 to 2500 km portions at each end of the loop are taken to be the chromospheric part of the loop, with the chromospheric model given by \citet{avrett_models_2008}. The loop resonant frequencies are taken from calculations of the Alfv\'en wave propagation and identifying the frequency at which wave transmission from the chromosphere into the corona is maximized.

We estimated the resonance angular frequency of the emerging loops to be $\Omega_{\text{EL resonance}}$ to be 0.351 $\text{rad~s}^{-1}$. We then run the model simulations assuming the fractionation process is driven by Alfv\'en waves at this frequency in both the emerging and the preexisting loops. This means at resonance with the emerging loops, and off resonance with the preexisting ones. A discussion on why this is likely to be an appropriate choice is provided in Section \ref{Wave Origin Discussion}.

\begin{figure*}
    \centering
    \includegraphics[width=1.0\textwidth]{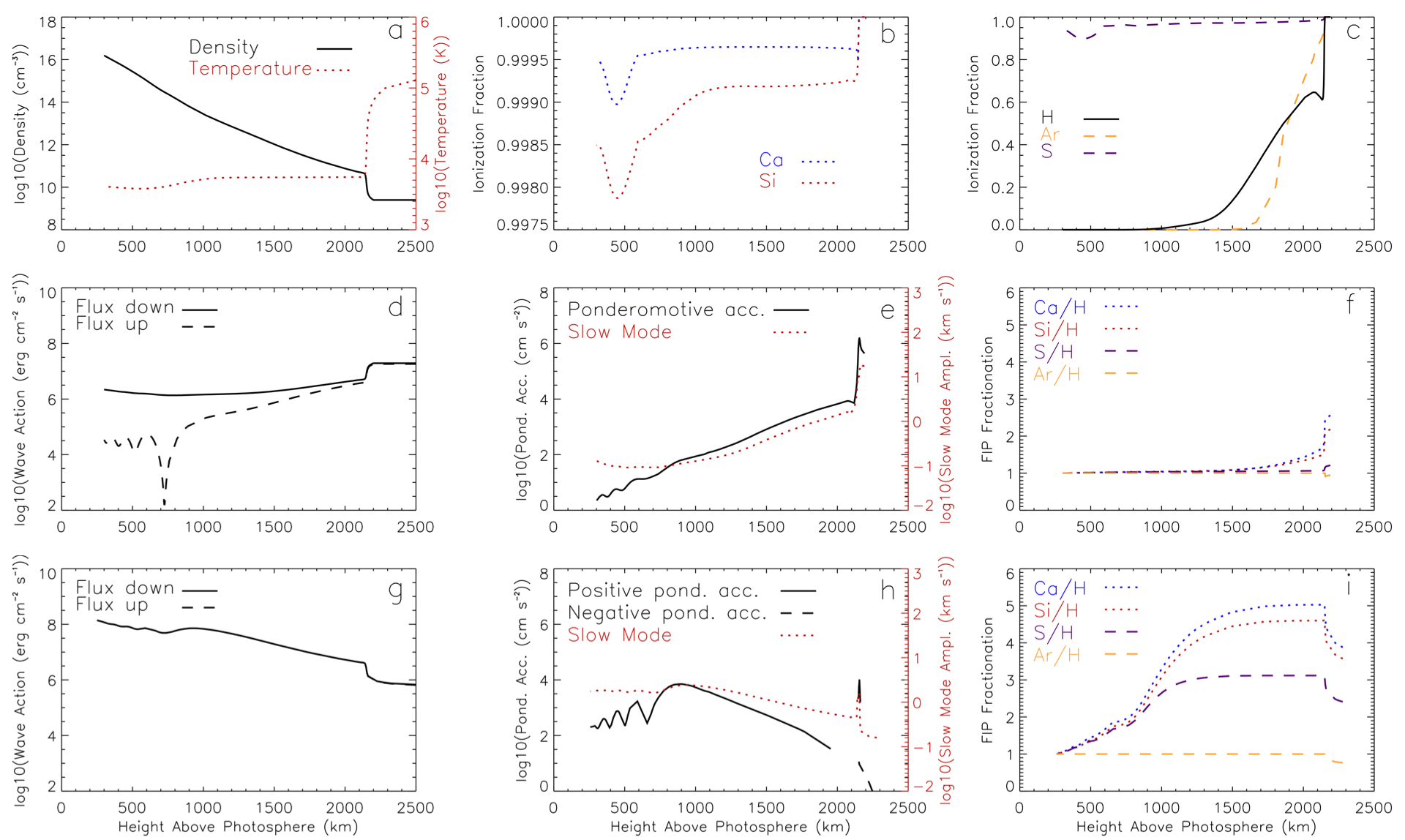}
    \caption{Ponderomotive force model predictions. First row shows the variation with height of the following parameters: a) electron temperature and density, b) ionization fraction variation for low-FIP elements, c) ionization fraction for high-FIP elements. Second row shows the variation with height of d) downward and upward Alfv\'en wave energy fluxes for waves of assumed coronal origin e) ponderomotive acceleration f) FIP bias relative to H in the resonant fractionation case. Third row shows the variation with height of g)  Alfv\'en wave energy fluxes; the downward and upward fluxes are identical, h) ponderomotive acceleration i) FIP bias relative to H in the non-resonant fractionation case.}
    \label{PF Model}
\end{figure*}

Results for the resonant case are shown in Figure \ref{PF Model}d,e,f. In this case, the ponderomotive acceleration starts to increase in the middle of the chromosphere and is highest at the top of the chromosphere and the transition region (Figure \ref{PF Model}e). As a result, the abundances of Si and Ca (relative to H, i.e. absolute abundances) start increasing slightly from the middle of the chromosphere, and show the highest enhancement around the top of the chromosphere and the transition region (Figure \ref{PF Model}f). We focus on the relative abundance predictions at the top of the transition region (i.e. 2500 km above the photosphere in these simulations) since the model suggests that abundance ratios are locked in once the plasma leaves this layer and is transported into the corona. Hence these are the ones to be compared with coronal observations. At the top of the transition region, Ar and S are essentially not fractionated. Ca and Si are both significantly fractionated. Ca shows a stronger fractionation than Si (Figure \ref{PF Model}f) which is likely because the ionization fraction of Ca is higher than the one of Si (Figure \ref{PF Model}b) which means a larger fraction of the Ca atoms are affected by the ponderomotive force. This could explain why the Hinode EIS observations show slightly higher Ca/Ar FIP bias values than Si/S FIP bias values in the emerging loops.

Results for the non-resonant case are shown in Figure \ref{PF Model}g,h,i. In this case, the fractionation happens lower down in the chromosphere and is overall stronger: the ponderomotive acceleration starts increasing at the bottom of the chromosphere, reaches a maximum at the middle of the chromosphere and then decreases (Figure \ref{PF Model}h). The abundances of Si and Ca (relative to H) start increasing from the bottom of the chromosphere, which results in stronger enhancements at the top of the transition region compared to the resonant case. Interestingly, in this case S behaves similarly to Si and Ca rather than Ar: it becomes enhanced as well, but to a lower degree than Si and Ca. As in the resonant case, Ar shows no fractionation. This could explain the strong difference between the Ca/Ar FIP bias and the Si/S FIP bias in the Hinode EIS measurements of the preexisting loops. As in the resonant case, Ca is more strongly enhanced than Si due to its higher ionization fraction. However, in the non-resonant case, the largest discrepancy between diagnostics comes from the fact that S experiences significant fractionation as well, i.e. does not behave like a high-FIP element anymore. Therefore, the enhancement of Si is underestimated when measured relative to S. 

\begin{deluxetable}{lcc}
\tabletypesize{\footnotesize}
\centering
\tablecolumns{5}
\tablehead{\colhead{Parameter} & \colhead{Emerging Loops} & \colhead{Preexisting Loops}}
\tablecaption{Parameters used for Ponderomotive Force Model Predictions} \label{PF_parameters}
\startdata
L & 100 Mm & 510 Mm\\
$\rho$ & $10^{9.5}\text{cm}^{-3}$ & $10^{8.5}\text{cm}^{-3}$\\
B & 250 G & 200G\\
$v_A$ & $10^4$ km s$^{-1}$& $2.5\times 10^4$ km s$^{-1}$\\
$f_{\text{resonance}}$ & 0.05 s$^{-1}$& 0.025 s$^{-1}$\\
$\Omega_{\text{resonance}}$ & 0.351 rad s$^{-1}$& 0.157 rad s$^{-1}$\\
$A_{\text{chromo}}$  & 0.22 km/s & 0.03 km/s\\
$A_{\text{coronal}}$ & 44 km/s & 14 km/s
\enddata
\tablecomments{Parameters listed here follow the same notation in Equations \ref{eq_f_resonance} and \ref{eq_alfven_speed}. For the wave amplitudes, $A_{\text{chromo}}$ is the chromospheric wave amplitude at the $\beta=1$ layer used as input for the model simulations and $A_{\text{coronal}}$ is the coronal wave amplitude predicted by the model.}
\end{deluxetable}

It is important to note that the main free parameter when running the model and making FIP bias predictions is the chromospheric amplitude $A_{\text{chromo}}$ of the wave that drives the fractionation process. In the absence of observations that can help constrain the amplitudes of these waves at the chromospheric level, we need to make a guess for the amplitudes to be able to make a FIP bias prediction. Small changes in the wave amplitudes result in large changes in the estimated fractionation strength at the top of the transition region.
However, while the strength of the fractionation is strongly dependent on the amplitude, the fractionation patterns for different elements mainly depend on the frequency of the wave driver (and, more specifically, on how close the frequency of the driver is to the resonant frequency of the loop) rather than the amplitude, so we can use the model predictions to obtain a qualitative understanding of the relative enhancement of different elements. 

In the example described in this section and shown in Figure \ref{PF Model}, we initiate the calculations with the wave amplitudes $A_{\text{chromo}}$ given in Table \ref{PF_parameters}. These are the input chromospheric wave amplitudes which resulted in fractionation patterns that best matched the Hinode EIS observations at 01:08 UT. The model predicts the waves to develop amplitudes $A_{\text{coronal}}$ in the coronal portions of the loops. In the chromosphere, where the FIP fractionation is calculated, the resonant wave amplitude returns to $A_{\text{chromo}}= 0.22$ km s$^{-1}$. The non-resonant wave amplitude is much larger, of order 1 km s$^{-1}$. This is characteristic of the non-resonant waves, in that more wave energy accumulates in the chromosphere than in the corona. 

Using the inputs listed in Table \ref{PF_parameters}, the model predicts the following results: in the emerging loops (resonant wave driver case), a Si/S FIP bias of 1.8 and Ca/Ar FIP bias of 2.7 and, in the preexisting loops (non-resonant wave driver case), a Si/S FIP bias of 1.5 and Ca/Ar FIP bias of 5.1. These values are directly calculated from the simulation composition patterns at transition region level shown in Figure \ref{PF Model}f,i. Following the same approach, $A_{\text{chromo}}$ can be changed slightly to obtain FIP bias values that match the Hinode EIS observations at 14:15 UT as well (see Table \ref{PF_simulations}). The key result, however, is that, while the predicted FIP bias values depend on the selected $A_{\text{chromo}}$, the significant differences between the two diagnostics depend on whether the wave is resonant or not.

\begin{deluxetable}{lcclcc}
\tabletypesize{\footnotesize}
\centering
\tablecolumns{5}
\tablehead{\multicolumn{3}{c}{Emerging Loops} & \multicolumn{3}{c}{Preexisting Loops}}
\tablecaption{Ponderomotive Force Model Predictions with Variable  $A_{\text{chromo}}$} \label{PF_simulations}
\startdata
$A_{\rm chromo}$ & Si/S & Ca/Ar& $A_{\rm chromo}$ & Si/S & Ca/Ar\\
0.20& 1.61& 2.31& 0.025& 1.32& 3.13\\
0.22& 1.80& 2.70& 0.03& 1.48& 5.1\\
0.26& 2.22& 4.00& 0.035& 1.68& 7.0\\
0.30& 2.87& 6.1&  0.04& 1.93& 14.7\\
\enddata
\tablecomments{For the wave amplitudes, $A_{\text{chromo}}$ is the chromospheric wave amplitude at the $\beta=1$ layer. $A_{\rm corona}$ changes in proportion. Parameters are chosen to span the range of Si/S and Ca/Ar in Table 2.}
\end{deluxetable}

While one could model Si/S and Ca/Ar coming from different strands with different wave populations within each loop, we consider it a success that the same fractionation process (i.e. one wave in each loop, of the same frequency) for both Si/S and Ca/Ar reduces the former and increases the latter in going from emerging to preexisting loops. One might achieve better agreement between model and observation with more waves, but this would be at the expense of more model parameters.

\subsection{Alfv\'en Waves Origin Discussion}
\label{Wave Origin Discussion}
The ponderomotive force model simulations suggest that the fractionation pattern observed in the two loop populations could be explained if the driver is resonant waves in the emerging loops and non-resonant waves in the preexisting loops. This naturally raises a question regarding the origin of these waves. We propose that the waves giving rise to the fractionation seen in the emerging loops have a coronal driver, and consider both a coronal and a photospheric driver for the for the waves giving rise to the fractionation observed in the preexisting loops. The emerging loops make up the very active core of the active region, where nanoflares, for example, can give rise to coronal Alfv\'en waves that are naturally at resonance with the loop. The preexisting loops are less active, which means an external driver is more likely to generate the waves needed for the fractionation.

The first candidate is of coronal origin. We speculate that resonant waves in the emerging loops could be communicated to the pre-existing loops where they will be non-resonant. To see this we write an equation of motion for waves on the pre-existing loop (2) forced by kink oscillations of the emerging loop (1) with displacement $x_2$
\begin{equation}
    \rho _2\left(\ddot{x_2} +\Omega _2^2x_2\right)=-{\partial\over\partial r}\left({\delta B_{\perp 1}^2\over 8\pi} +{B_0\delta B_{||1}\over 4\pi}\right).
\end{equation}
Here we assume $\delta B_1\propto 1/r$, where $\delta B_1$ is the magnetic field perturbation in emerging loop 1, with components perpendicular and parallel to the ambient magnetic field $B_0$ indicated. The first Alfv\'enic term on the right hand side $\delta B_{\perp 1}^2/8\pi = \rho _1\delta v_{\perp 1}^2/2$ and and oscillates at $2\Omega _1$ to give
\begin{equation}
    {x_2\over x_1} = {\rho_1\over\rho_2}{2\Omega _1^2\over\Omega _2^2-4\Omega _1^2}{x_1\over R}.
\end{equation}
where $R$ is the separation between the flux tubes.
With $x_1=44/0.351 = 125$ km, equation 4 indicates that $x_2/x_1=0.3$ requires a separation between flux tubes of 2" - 4", much smaller than the observed separation of order 100''. 

The effect of the second, compressive, term on the right hand side is more model dependent
\citep[e.g.][]{mikhalyaev_oscillations_2005,verwichte_fast_2006}. The $r$-component of the wavevector exterior to the emerging loop can be written when the plasma beta $\beta << 1$
\begin{equation}
    k_r^2=\Omega ^2-k_z^2v_{Ae}^2 = {v_{A1}^2-v_{Ae}^2\over 1+\rho _e/\rho _1}
\end{equation}
where $\Omega ^2=\left(B_e^2+B_0^2\right)/\left(\rho _e+\rho _1\right)/4\pi $ is the tube oscillation frequency in terms of exterior and interior ambient magnetic fields $B_e$, $B_0$, and densities $\rho _e$, $\rho _1$ \citep{mikhalyaev_oscillations_2005}. When $v_{Ae} > v_{A1}$ (the usual case), $k_r$ is imaginary and the exterior wave is evanescent, meaning that the kink oscillations are trapped inside the loop, making transfer of wave energy from one loop to the other unlikely. Such a situation has been studied in detail for magnetosonic waves escaping from reconnection current sheets by \citet{provornikova_reflection_2018}.The forgoing is doubtless oversimplified, and many possibilities must exist for the excitation of wave modes as magnetic flux emerges. Waves generated by nanoflares are, however, highly localised to single loops rather than loop populations which raises questions about whether this type of wave transfer could realistically take place. 

A second candidate could be of photospheric origin, and include p- and g-mode oscillations or other perturbations of the photospheric plasma flows generated by the flux emergence happening in the close vicinity of the preexisting loops footpoints (see Figure \ref{Continuum_evolution}). These can act as a driver for both resonant and non-resonant waves. Among the wide range of possible photospheric perturbations, those with long wavelengths could naturally couple and perturb the neighbouring preexisting loops population at the same time. In this regard, it is worth noting that the global rotational motion seen in the active region suggests a photospheric or sub-photospheric driver on large spatial scales. Interestingly, \citet{grant_propagation_2022} have detected coherent waves across multiple pores in the photosphere suggesting a coupled wave excitation mechanism and a driver acting on scales of several tens of Mm. The rotational motion observed in the active region studied here could also drive torsional Alfv\'en waves. The associated spatial scales of the driver may explain coupled behaviour of different loops. The S enhancement in the preexisting loops (as suggested by the Hinode EIS observations) and the fact that, as suggested by the model, the fractionation process takes place lower down in the chromosphere in the non-resonant case support the scenario of a photospheric origin for the waves driving fractionation in the preexisting loops. It is, however, important to note that the low frequencies and long wavelengths implied for 3 or 5 minutes p-modes which make them reflect more easily in the chromosphere mean that high chromospheric wave amplitudes, $> 10$ km s$^{-1}$ are required. Higher frequency non-resonant waves, such as modeled in Figure 6, reduce this amplitude to $\sim 1$ km s$^{-1}$.

These are a couple of options that could explain the presence of non-resonant waves in the preexisting loop. However, understanding the exact origin of these waves is beyond the scope of this work and should be investigated in depth in a separate study.

\section{Summary and Discussion}
\label{Discussion}
Spectral analysis of Hinode EIS observations of NOAA AR 12665 show very different FIP bias values in two parts of the active region. The emerging loops, i.e. the new part of the active region, show enhanced Si/S FIP bias (2.3 to 3.0) and slightly higher Ca/Ar FIP bias (2.6 to 3.3). The preexisting loops, i.e. the old part of the active region, show more modest Si/S FIP bias (1.8 to 2.0) but very strong Ca/Ar FIP bias (4.3 to 7.8). The Ca/Ar FIP bias is slightly higher than the Si/S FIP bias in the emerging loops, but much higher than the Si/S FIP in the preexisting loops. We find that the ponderomotive force model is able to predict this effect using simple assumptions about the properties of the waves driving the fractionation process in the two loops.

We propose that the fractionation pattern observed in the emerging loops can be given by resonant Alfv\'en waves of coronal origin. In this case, Ar and S show no fractionation, while Ca and Si show significant fractionation (Ca slightly higher than Si which could explain the slightly stronger Ca/Ar FIP bias compared to Si/S FIP bias). This can be explained by fractionation occurring at the top of the chromosphere \citep[as was previously suggested for the hot core loops from measurements of significantly higher FIP bias values by][]{brooks_source_2021}. 

The fractionation pattern observed in the preexisting loops can be given by non-resonant waves. In this case, Ca and Si show stronger fractionation than in the previous case (Ca again slightly higher than Si) and Ar again shows no fractionation. The key difference is that, in these conditions, S shows significant fractionation as well, resulting in a much lower Si/S FIP bias than Ca/Ar FIP bias. This can be explained by fractionation occurring lower down in the chromosphere \citep[as was previously suggested by][]{laming_element_2019}. Note that although the two diagnostics measure different FIP bias levels, they are both detecting coronal abundances in both the hot core loops and pre-existing loops. It is the combination of Si/S and Ca/Ar FIP bias measurements that allow further probing of the model predictions, and the development of the resonant/non-resonant wave explanation of the wider loop environment. 

It is important to note that the model is only analysing the chromospheric and transition region environment, predicting abundances at the top of the transition region. Transport mechanisms from the top of the transition region to the corona must be considered to be able to make a prediction of the FIP bias in the corona. The model suggests that the abundance change at the top of the transition region can be reached within minutes \citep{laming_fip_2015}. However, previous studies \citep{widing_rate_2001, baker_coronal_2018}, found that the FIP bias increases with time in the emergence phase over hours to days. This indicates that transport processes that bring the fractionated plasma into the corona are much slower than the processes that drive the fractionation, so predicted chromospheric abundances will not be reflected in the corona right away. Nevertheless, assuming that these coronal transport timescales are similar in the emerging and preexisting loops, we can compare qualitative trends predicted by the ponderomotive force model to our Hinode EIS observations.

We observe an increase in both the Si/S FIP bias and the Ca/Ar FIP bias of the emerging loops over the 13 hour period between the two Hinode EIS scans. This could indicate that the fractionated plasma is slowly being transported to the corona, i.e. the fractionation pattern at the top of the transition region is slowly being reflected in the corona. The strong increase of the Ca/Ar FIP bias in the preexisting loops over the 13 hour period could be explained following the same reasoning, with the exception that the final Ca/Ar FIP bias is higher so the increase appears to be more drastic. The steady increase assumption requires reasonably quiet coronal conditions. However, we observe 1 M-class and 4 C-class flares happening at the boundary between these two loop populations in between the first and the second Hinode EIS scans. In the EUV, previous studies showed that flaring can either temporarily reduce the FIP bias to photospheric values \citep{warren_measurements_2014} or increase it \citep{to_evolution_2021}. While the boxes selected for our FIP bias measurements are located further away from the flaring sites, we cannot exclude the possibility that the flaring activity influences the coronal transport mechanisms in the loops we are studying.

Finally, this result is particularly relevant for connection science studies. Typically the FIP bias diagnostics used in remote sensing studies (e.g., Si/S, Fe/S, Ca/Ar, Mg/Ne) are different from the ones used for in-situ measurements (e.g., Fe/O), so understanding when these diagnostics behave differently is important for connnecting the two types of measurements. According to the ponderomotive force model, resonant waves drive little fractionation in larger loops (e.g., the ones in this study) and no fractionation in open loops \citep{laming_fip_2015}. However, as the present work suggests, fractionation can still be driven in these loops by non-resonant waves if an external driver is present. In this case the Si/S FIP bias would be very low (and this can be extended to other low-FIP elements relative to S) so the area under study could mistakenly be believed to show no/weak fractionation unless a second diagnostic (not including S) is used. Therefore, understanding what differences to expect between different FIP bias diagnostics and being able to predict in what conditions S starts experiencing significant fractionation is important for connecting in-situ plasma parcels to their origin on the Sun. This is especially important given that S is commonly used in both remote sensing and in-situ FIP bias diagnostics.


\begin{acknowledgments}
We would like to thank Tom Van Doorsselaere for insightful comments and discussions that helped develop some of the ideas presented in this work.
The EIS Chief Observer for July 9 was H.~P.~Warren. 
T.M. acknowledges funding the Science and Technology Funding Council (STFC) PhD studentship ST/V507155/1 and the support of a Royal Astronomical Society Field Trip grant to carry out part of this work.
The work of D.H.B. was performed under contract to the Naval Research Laboratory and was funded by the NASA Hinode program.
J.M.L. is supported by NASA HSR Grant NNH22OB102 and by Basic Research Funds of the Office of Naval Research.
D.B. is funded under Solar Orbiter EUI Operations grant number ST/X002012/1 and Hinode Ops Continuation 2022-25 grant number ST/X002063/1.
A.W.J. acknowledges funding from the STFC Consolidated Grant ST/W001004/1.
D.M.L. is grateful to the STFC for the award of an Ernest Rutherford Fellowship (ST/R003246/1).
L.v.D.G. acknowledges the Hungarian National Research, Development and Innovation Office grant OTKA K-131508.
Hinode is a Japanese mission developed and launched by ISAS/JAXA, collaborating with NAOJ as a domestic partner, and NASA and STFC (UK) as international partners. Scientific operation of Hinode is performed by the Hinode science team organized at ISAS/JAXA. Support for the post-launch operation is provided by JAXA and NAOJ (Japan), STFC (UK), NASA, ESA, and NSC (Norway). 
CHIANTI is a collaborative project involving George Mason University, the University of Michigan (USA), University of Cambridge (UK) and NASA Goddard Space Flight Center (USA).
\end{acknowledgments}

%

\software{SolarSoftWare \citep{freeland_data_1998}, EISPAC \citep{weberg_eispac_2023}}






\bibliography{references}{}
\bibliographystyle{aasjournal}



\end{document}